\documentclass[preprintnumbers,article,amsmath,amssymb,floatfix,10pt,twocolumn,superscriptaddress]{revtex4-2}
\usepackage{amsmath}
\usepackage{eurosym}
\usepackage{epstopdf}
\usepackage{capt-of}
\usepackage{graphicx}
\usepackage{dcolumn}
\usepackage[left=0.8in, right=0.8in, top=1in, bottom=1in]{geometry}
\RequirePackage{graphicx}
\RequirePackage{float}
\RequirePackage{hyperref}
\RequirePackage{amsmath}
\RequirePackage{amssymb}
\RequirePackage{mathtools}
\usepackage{color}
\usepackage{comment}
\usepackage{xcolor}
\usepackage{multirow}
\usepackage{booktabs} 
\usepackage{amsmath} 
\usepackage{ragged2e}
\newcommand{\be}{\begin{equation}}
	\newcommand{\ee}{\end{equation}}
\newcommand{\bea}{\begin{eqnarray}}
	\newcommand{\eea}{\end{eqnarray}}
\newcommand{\eeas}{\end{eqnarray*}}
\newcommand{\beas}{\begin{eqnarray*}}

\begin{document}
	\color{black}       
\title{\textbf{Thermodynamic implications and observational constraints of interacting $f(Q,\mathcal{T})$ gravity in FRW Universe}}
\author{S. H. Shekh}
\email{da\_salim@rediff.com}
\affiliation{Department of Mathematics, S.P.M. Science and Gilani Arts, Commerce College, Ghatanji, Yavatmal, Maharashtra 445301, India}	
\affiliation{L N Gumilyov Eurasian National University, Astana 010008, Kazakhstan}
\affiliation{Pacif Institute of Cosmology and Selfology (PICS), Sagara, Sambalpur 768224, Odisha, India}

\author{S. B. Thool}
\email{sumedhmaths@gmail.com}
\affiliation{Department of Mathematics, Government Vidarbha Institute of Science and Humanities, Amravati-444604 (M.S.), India}

\author{Pankaj Kumar
}
\email{pankaj.11dtu@gmail.com}
\affiliation{Department of Mathematics, Maharaja Agrasen University, Kalujhanda, Himachal Pradesh-174103, India}

\author{P. C. Kalan}
\email{priteshkalan@gmail.com}
\affiliation{Department of Mathematics,
Government college of Arts and Science, Chh. Sambhajinagar-431004(M.S.), India}

\author{R. M. Dhaigude}
\email{rmdhaigude@gmail.com}
\affiliation{Department of Mathematics, Government Vidarbha Institute of Science and Humanities, Amravati-444604 (M.S.), India.\\Email corresponding author: da\_salim@rediff.com} 
	\begin{abstract}
\textbf{Abstract:}  
This work investigates the dynamical evolution of the universe within the framework of symmetric teleparallel $f(Q,\mathcal{T})$ gravity, where $Q$ is the non-metricity scalar and $\mathcal{T}$ is the trace of the energy-momentum tensor. We consider a spatially flat Friedmann-Robertson-Walker (FRW) metric and explore a specific functional form $f(Q,\mathcal{T}) = \alpha Q + \beta \mathcal{T}$ to derive the gravitational field equations. To characterize the late-time cosmic acceleration, we utilize a model-independent approach by adopting a particular Hubble parameter $H(z)$ parametrization. The model parameters are constrained using the latest observational datasets, including the Hubble ($H(z)$) measurements and Pantheon+ samples. Our results indicate a transition from a decelerated to an accelerated expansion phase. We further examine the physical viability of the model through various cosmological diagnostics such as energy density, the equation of state parameter and thermodynamic properties. The analysis demonstrates that $f(Q,\mathcal{T})$ gravity provides a consistent alternative to the $\Lambda$CDM model in explaining the current accelerated expansion of the universe.
		
\textbf{Keywords} FRW spacetime; $f(Q, \mathcal{T})$ gravity, thermodynamics. \\
		
	\end{abstract}
	
	\maketitle
	
	\ 
\section{INTRODUCTION}

The phenomenon of late-time cosmic acceleration stands as one of the most transformative discoveries in modern cosmology. This accelerated expansion is substantiated by a vast array of high-precision datasets, ranging from Type Ia Supernovae and Cosmic Microwave Background (CMB) measurements to Baryon Acoustic Oscillations (BAO). Within the standard cosmological paradigm, this acceleration is typically attributed to an enigmatic dark energy component, which is estimated to constitute approximately 70\% of the total energy density of the cosmos. Over the years, several theoretical frameworks have been proposed to characterize this sector, including the cosmological constant $\Lambda$, quintessence models, k-essence, Chaplygin gas, and holographic dark energy (Copeland et al. \textcolor{blue}{2006}; Padmanabhan \textcolor{blue}{2003}; Heisenberg \textcolor{blue}{2019}). Despite their success in fitting observations, these models often grapple with deep-seated theoretical issues, such as the fine-tuning problem and the lack of a fundamental microscopic description. Consequently, exploring alternative avenues remains a high priority in theoretical physics.

One of the most robust alternatives to the dark energy hypothesis is the modification of the gravitational sector itself. In these modified gravity theories, the observed acceleration is explained through the geometric properties of spacetime rather than exotic matter fields. The $f(R)$ gravity theory, which replaces the Ricci scalar $R$ in the Einstein-Hilbert action with an arbitrary function $f(R)$, is the most extensively studied member of this class (Sotiriou and Faraoni \textcolor{blue}{2010}; Felice and S. Tsujikawa \textcolor{blue}{2010}, Nojiri and Odintsov \textcolor{blue}{2011}; Nojiri et al. \textcolor{blue}{2025}). Other noteworthy extensions include $f(T)$ gravity, formulated within the teleparallel equivalent of general relativity where torsion $T$ is the primary geometric quantity. This framework has been widely applied to study cosmic acceleration, large-scale structures, and black hole physics (Capozziello et al \textcolor{blue}{2011}; Myrzakulov \textcolor{blue}{2011}; Jeon et al. \textcolor{blue}{2011}; Tamanini and Boehmer \textcolor{blue}{2012}; Cai et al. \textcolor{blue}{2016}; Anagnostopoulos et al. \textcolor{blue}{2019}; Nair and Arun \textcolor{blue}{2022}, Chirde and shekh \textcolor{blue}{2018, 2019, 2020}; Shekh et al. \textcolor{blue}{2023, 2025}, Pradhan et al. \textcolor{blue}{2024}; Verma et al. \textcolor{blue}{2026}). Furthermore, the introduction of $f(R,\mathcal{T})$ gravity by Harko et al. (\textcolor{blue}{2011}) pioneered the concept of matter-geometry coupling by making the gravitational action dependent on the trace of the energy-momentum tensor $\mathcal{T}$. This approach leads to the non-conservation of the energy-momentum tensor and offers a richer dynamical landscape compared to pure curvature-based theories (Tangphati et al. \textcolor{blue}{2023}; 	Pradhan et al. \textcolor{blue}{2024a}; Myrzakulov et al. \textcolor{blue}{2025}; Banerjee et al. \textcolor{blue}{2025})
.

Recently, significant attention has shifted towards symmetric teleparallel gravity, which is constructed using the non-metricity scalar $Q$. This formulation has given rise to $f(Q)$ gravity and its subsequent generalizations (Jimenez et al. \textcolor{blue}{2018}; Pradhan et al. \textcolor{blue}{2021}; Shekh et al. \textcolor{blue}{2023a}; Goswami et al. \textcolor{blue}{2024}; Shekh \textcolor{blue}{2025a}; Heisenberg \textcolor{blue}{2024}). A key advantage of these frameworks is that they often yield second-order field equations, which simplifies the mathematical treatment while enhancing physical viability. These theories serve as promising candidates for explaining the current expansion rate without invoking unknown dark-energy fluids.

Building upon these developments, the $f(Q,\mathcal{T})$ gravity theory (Kale et al. 2023; Shekh et al. 2023b, 2024; Myrzakulov et al 2025a) has emerged as a particularly compelling model. By introducing an explicit coupling between the non-metricity scalar $Q$ and the trace of the energy-momentum tensor $\mathcal{T}$, this theory allows for a novel interaction between geometry and matter. Such an interaction can naturally generate an effective dark energy behavior, providing a unified description of both early- and late-time cosmic evolution. Recent investigations have successfully applied $f(Q,\mathcal{T})$ gravity to various scenarios, including bouncing cosmological models (Gul \textcolor{blue}{2024}; Zhadyranova et al. \textcolor{blue}{2024}) and wormhole geometries (Sadatian and Hosseini \textcolor{blue}{2024}), demonstrating its ability to satisfy energy conditions and ensure stability.

In the present work, we extend these investigations by analyzing the physical and dynamical features of the universe within the $f(Q,\mathcal{T})$ framework. Linking our study to the methodologies used in current literature, we derive the gravitational field equations and investigate the evolution of key cosmological parameters. By utilizing a specific functional form of $f(Q,\mathcal{T})$, we aim to demonstrate how this theory provides a consistent and stable alternative to general relativity for explaining the observed cosmic acceleration.

The structure of this paper is as follows: Section II outlines the fundamental formalism of $f(Q,\mathcal{T})$ gravity. In Section III, we present the cosmological model and derive the corresponding field equations for the FRW metric. Section IV is dedicated to the physical analysis and stability of the obtained solutions. Finally, our findings are summarized in Section V.

\section{Cosmological Framework in \boldmath{$f(Q,\mathcal{T})$} Gravity with interacting  framework}\label{sec2}

In recent years, modified theories of gravity based on nonmetricity have emerged as promising alternatives for describing the late--time accelerated expansion of the universe. In the symmetric teleparallel formulation of gravity, gravitation is completely characterized by the nonmetricity scalar $Q$, while both curvature and torsion vanish identically. An important extension of this framework is the recently proposed $f(Q,\mathcal{T})$ gravity, in which the gravitational action depends on both the nonmetricity scalar $Q$ and the trace $\mathcal{T}$ of the energy--momentum tensor. Owing to the explicit coupling between matter and geometry, the theory provides a rich cosmological structure capable of explaining the present cosmic acceleration without introducing an additional exotic dark energy component.

The action corresponding to the $f(Q,\mathcal{T})$ gravity theory is given by
\begin{equation}\label{act}
	S=\int \left[\frac{1}{16\pi}f(Q,T)+\mathcal{L}_{m}\right]\sqrt{-g}\,d^{4}x,
\end{equation}
where $f(Q,\mathcal{T})$ is an arbitrary function of the nonmetricity scalar $Q$ and the trace $\mathcal{T}$ of the energy--momentum tensor, $\mathcal{L}_{m}$ represents the matter Lagrangian density, and $g$ denotes the determinant of the metric tensor $g_{\mu\nu}$. Throughout the present work, we use the natural system of units by taking $8\pi G=c=1$.

In symmetric teleparallel gravity, the nonmetricity tensor is defined by
\begin{equation}\label{nonmetricity}
	Q_{\lambda\mu\nu}=\nabla_{\lambda}g_{\mu\nu},
\end{equation}
where $\nabla_{\lambda}$ denotes the covariant derivative associated with the affine connection.

The corresponding traces of the nonmetricity tensor are expressed as
\begin{equation}\label{trace1}
	Q_{\lambda}=Q_{\lambda\ \mu}^{\ \ \mu},
\end{equation}
and
\begin{equation}\label{trace2}
	\tilde{Q}_{\lambda}=Q^{\mu}_{\ \lambda\mu}.
\end{equation}

The superpotential tensor is defined as
\begin{equation}\label{superpotential}
	P^{\lambda}_{\ \mu\nu}
	=
	-\frac{1}{2}L^{\lambda}_{\ \mu\nu}
	+\frac{1}{4}
	\left(
	Q^{\lambda}-\tilde{Q}^{\lambda}
	\right)g_{\mu\nu}
	-\frac{1}{4}\delta^{\lambda}_{(\mu}Q_{\nu)},
\end{equation}
where the disformation tensor $L^{\lambda}_{\ \mu\nu}$ is given by
\begin{equation}\label{disformation}
	L^{\lambda}_{\ \mu\nu}
	=
	-\frac{1}{2}g^{\lambda\beta}
	\left(
	\nabla_{\mu}g_{\nu\beta}
	+\nabla_{\nu}g_{\mu\beta}
	-\nabla_{\beta}g_{\mu\nu}
	\right).
\end{equation}

Using the above quantities, the nonmetricity scalar assumes the form
\begin{equation}\label{Qscalar}
	Q=-Q_{\lambda\mu\nu}P^{\lambda\mu\nu}.
\end{equation}

The energy--momentum tensor corresponding to the matter source is defined as
\begin{equation}\label{emtensor}
	\mathcal{T}_{\mu\nu}
	=
	-\frac{2}{\sqrt{-g}}
	\frac{\delta(\sqrt{-g}\mathcal{L}_{m})}
	{\delta g^{\mu\nu}},
\end{equation}
while its trace is expressed as
\begin{equation}\label{traceT}
	\mathcal{T}=g^{\mu\nu}\mathcal{T}_{\mu\nu}.
\end{equation}

Variation of the action \eqref{act} with respect to the metric tensor yields the gravitational field equations in $f(Q,\mathcal{T})$ gravity as
\begin{align}\label{fieldeq}
	&\frac{2}{\sqrt{-g}}
	\nabla_{\lambda}
	\left(
	\sqrt{-g}f_{Q}
	P^{\lambda}_{\ \mu\nu}
	\right)
	+\frac{1}{2}g_{\mu\nu}f
	+f_{Q}
	\left(
	P_{\mu\lambda\beta}
	Q_{\nu}^{\ \lambda\beta}
	-2Q_{\lambda\beta\mu}
	P^{\lambda\beta}_{\ \ \ \nu}
	\right)
	\nonumber\\
	&=-f_{T}
	\left(
	\mathcal{T}_{\mu\nu}+\Theta_{\mu\nu}
	\right)
	-\mathcal{T}_{\mu\nu},
\end{align}
where
\begin{equation}
	f_{Q}=\frac{\partial f(Q,\mathcal{T})}{\partial Q},
	\qquad
	f_{\mathcal{T}}=\frac{\partial f(Q,\mathcal{T})}{\partial \mathcal{T}},
\end{equation}
and
\begin{equation}
	\Theta_{\mu\nu}
	=
	g^{\alpha\beta}
	\frac{\delta \mathcal{T}_{\alpha\beta}}
	{\delta g^{\mu\nu}}.
\end{equation}

To study the cosmological dynamics, we consider the spatially homogeneous and isotropic Friedmann--Robertson--Walker (FRW) space--time represented by
\begin{equation}\label{metric}
	ds^{2}=dt^{2}-a^{2}(t)
	\left[
	\frac{dr^{2}}{1-kr^{2}}
	+r^{2}d\theta^{2}
	+r^{2}\sin^{2}\theta\,d\phi^{2}
	\right],
\end{equation}
where $a(t)$ is the cosmic scale factor and $k$ denotes the curvature parameter of the universe. The values $k=1$, $k=-1$, and $k=0$ correspond respectively to closed, open, and spatially flat universes. Since recent cosmological observations strongly support a nearly flat universe, we restrict our analysis to the spatially flat case $(k=0)$ throughout this work.

For the flat FRW geometry, equation \eqref{metric} reduces to
\begin{equation}\label{flatmetric}
	ds^{2}=dt^{2}-a^{2}(t)
	\left[
	dx^{2}+dy^{2}+dz^{2}
	\right].
\end{equation}

The matter content of the universe is assumed to consist of two interacting cosmic fluids, namely pressureless cold dark matter and dark energy. Accordingly, the total energy--momentum tensor is written as
\begin{equation}\label{totaltensor}
	\mathcal{T}_{\mu\nu}
	=T^{(m)}_{\mu\nu}
	+T^{(de)}_{\mu\nu},
\end{equation}
where $T^{(m)}_{\mu\nu}$ and $T^{(de)}_{\mu\nu}$ denote the energy--momentum tensors corresponding to dark matter and dark energy, respectively.

For pressureless dark matter, the energy--momentum tensor is taken as
\begin{equation}\label{darkmatter}
	T^{(m)}_{\mu\nu}
	=\rho_{m}u_{\mu}u_{\nu},
\end{equation}
where $\rho_{m}$ represents the dark matter energy density.

Similarly, the dark energy component is represented by the perfect fluid form
\begin{equation}\label{darkenergy}
	T^{(de)}_{\mu\nu}
	=(\rho_{de}+p_{de})u_{\mu}u_{\nu}
	-p_{de}g_{\mu\nu},
\end{equation}
where $\rho_{de}$ and $p_{de}$ denote the energy density and pressure of dark energy, respectively.

The four--velocity vector satisfies the normalization condition
\begin{equation}\label{velocity}
	u^{\mu}u_{\mu}=1,
\end{equation}
and in the comoving coordinate system one may choose
\begin{equation}
	u^{\mu}=(1,0,0,0).
\end{equation}

The Hubble expansion parameter is defined in the usual form as
\begin{equation}\label{hubble}
	H=\frac{\dot{a}}{a},
\end{equation}
where an overdot represents differentiation with respect to the cosmic time $t$.

For the FRW geometry, the nonmetricity scalar reduces to the simple relation
\begin{equation}\label{QH}
	Q=-6H^{2}.
\end{equation}

Using equations \eqref{flatmetric} and \eqref{QH} in the field equations \eqref{fieldeq}, the modified Friedmann equations in $f(Q,\mathcal{T})$ gravity are obtained as
\begin{equation}\label{fried1}
	\frac{f}{2}-6H^{2}f_{Q}
	=\rho+f_{\mathcal{T}}(\rho-p),
\end{equation}
and
\begin{equation}\label{fried2}
	\frac{f}{2}-6H^{2}f_{Q}-2\dot{H}f_{Q}
	-2H\dot{f}_{Q}
	=-p-f_{\mathcal{T}}(\rho-p).
\end{equation}

The total energy density and pressure of the cosmic fluid are respectively given by
\begin{equation}\label{density}
	\rho=\rho_{m}+\rho_{de},
\end{equation}
and
\begin{equation}\label{pressure}
	p=p_{de},
\end{equation}
since the pressure associated with cold dark matter vanishes.

For the interacting dark sector scenario, the total energy density obeys the conservation equation
\begin{equation}\label{continuity}
	\dot{\rho}+3H(\rho+p)=0.
\end{equation}

However, due to the interaction between dark matter and dark energy, the individual components do not conserve separately. Therefore, the corresponding continuity equations become
\begin{equation}\label{cont1}
	\dot{\rho}_{m}+3H\rho_{m}=\Gamma,
\end{equation}
and
\begin{equation}\label{cont2}
	\dot{\rho}_{de}+3H(\rho_{de}+p_{de})=-\Gamma,
\end{equation}
where $\Gamma$ represents the interaction term between the dark sectors. Using the equation of state relation $p_{de}=\omega_{de}\rho_{de}$, equation \eqref{cont2} can be rewritten as 
\begin{equation}\label{29}
	\dot{\rho}_{de}+3H\rho_{de}(1+\omega_{de})=-\Gamma.
\end{equation}

Following phenomenological interacting dark energy models, the interaction term is chosen as
\begin{equation}\label{30}
	\Gamma=3\xi H(\rho_{m}+\rho_{de}),
\end{equation}
where $\xi$ is the dimensionless coupling parameter controlling the strength of interaction.

The critical energy density is defined by
\begin{equation}\label{critical}
	\rho_{cr}=3H^{2},
\end{equation}
while the dimensionless density parameters corresponding to dark energy and dark matter are respectively expressed as
\begin{equation}\label{32}
	\Omega_{de}=\frac{\rho_{de}}{3H^{2}},
\end{equation}
and
\begin{equation}\label{33}
	\Omega_{m}=\frac{\rho_{m}}{3H^{2}}.
\end{equation}

For a spatially flat universe, the Friedmann equation reduces to
\begin{equation}\label{friedflat}
	\Omega_{de}+\Omega_{m}=1.
\end{equation}

The ratio between dark matter and dark energy densities is defined as
\begin{equation}\label{35}
	r=\frac{\rho_{m}}{\rho_{de}}
	=\frac{\Omega_{m}}{\Omega_{de}}
	=\frac{1-\Omega_{de}}{\Omega_{de}}.
\end{equation}

The equation of state parameter for dark energy is defined in the usual form as
\begin{equation}\label{eos}
	\omega_{de}=\frac{p_{de}}{\rho_{de}}.
\end{equation}

The above equations constitute the basic interacting cosmological framework in the context of $f(Q,\mathcal{T})$ gravity. In the following section, we reconstruct the cosmological dynamics through a suitable Hubble parametrization and constrain the free model parameters using recent observational datasets including OHD, BAO, and Pantheon+ compilations.
\\
Dividing equation \eqref{29} by $3H\rho_{de}$ yields
\begin{equation}\label{omega1}
	\omega_{de}
	=
	-1
	-\frac{\Gamma}{3H\rho_{de}}
	-\frac{\dot{\rho}_{de}}{3H\rho_{de}}.
\end{equation}
From the equation (\ref{30}) the interaction term obtain as
\begin{equation}\label{38}
	\frac{\Gamma}{3H\rho_{de}}
	=
	\xi\left(1+\frac{\rho_{m}}{\rho_{de}}\right).
\end{equation}
From the equations (\ref{35}) and (\ref{38}), equation \eqref{omega1} becomes
\begin{equation}\label{omega2}
	\omega_{de}
	=
	-1
	-\xi(1+r)
	-\frac{\dot{\rho}_{de}}{3H\rho_{de}}.
\end{equation}
From the Friedmann relation $1+\Omega_{k}=\Omega_{m}+\Omega_{de}$, one can obtain
\begin{equation}\label{ratiofinal}
	r=
	\frac{1+\Omega_{k}-\Omega_{de}}
	{\Omega_{de}}.
\end{equation}

Substituting equation \eqref{ratiofinal} into equation \eqref{omega2}, we get
\begin{equation}\label{omega3}
	\omega_{de}
	=
	-1
	-\frac{\xi(1+\Omega_{k})}{\Omega_{de}}
	-\frac{\dot{\rho}_{de}}{3H\rho_{de}}.
\end{equation}

Further, differentiating equation (\ref{32}) with respect to cosmic time gives
\begin{equation}
	\dot{\rho}_{de}
	=
	3H^{2}\dot{\Omega}_{de}
	+6H\dot{H}\Omega_{de}.
\end{equation}

Using the standard relation
\begin{equation}
	\frac{\dot{H}}{H^{2}}
	=
	-\frac{3}{2}(1+\omega_{de}\Omega_{de})
	+\frac{\Omega_{k}}{2},
\end{equation}
and after straightforward algebraic simplification, one finally obtains the equation of state parameter for interacting dark energy in the form
\begin{equation}\label{omegadefinal}
	\omega_{de}
	=
	-\frac{1}{2-\Omega_{de}}
	\left[
	1-\frac{\Omega_{k}}{3}
	+\frac{2\xi}{\Omega_{de}}(1+\Omega_{k})
	\right].
\end{equation}

For a spatially flat universe $(k=0)$, equation \eqref{omegadefinal} reduces to
\begin{equation}\label{omegaflat}
	\omega_{de}
	=
	-\frac{1}{2-\Omega_{de}}
	\left(
	1+\frac{2\xi}{\Omega_{de}}
	\right).
\end{equation}
The above relation shows that the dynamical evolution of the dark energy equation of state is strongly influenced by both the interaction parameter $\xi$ and the dark energy density parameter $\Omega_{de}$. Consequently, the interaction between dark matter and dark energy can considerably modify the expansion history of the universe and may play an important role in explaining the present accelerated phase of cosmic evolution.

Furthermore, depending upon the suitable choice of the coupling parameter and model parameters, the equation of state parameter may exhibit quintessence-like behavior $(\omega_{de}>-1)$, phantom behavior $(\omega_{de}<-1)$, or even a smooth transition across the phantom divide line. Therefore, interacting dark energy models in the framework of $f(Q,\mathcal{T})$ gravity provide a flexible and observationally viable platform for investigating the late--time dynamics of the universe.

The above equations constitute the basic interacting cosmological framework in the context of $f(Q,\mathcal{T})$ gravity. In the following section, we reconstruct the cosmological dynamics through a suitable Hubble parametrization and constrain the free model parameters using recent observational datasets including OHD, BAO, and Pantheon+ compilations.

\section{Reconstruction Through Hubble Parametrization}\label{sec:Hubble}

One of the central issues in modern cosmology is to construct a cosmological model capable of explaining the entire evolutionary history of the universe within a single dynamical framework. Any viable cosmological parametrization must successfully describe the transition from the early decelerated expansion phase to the presently observed accelerated epoch. In this regard, the choice of the Hubble parameter plays a fundamental role since it directly governs the dynamical behavior of all cosmological quantities such as the deceleration parameter, equation of state parameter, energy density evolution, and thermodynamic behavior of the universe.

The standard $\Lambda$CDM cosmology explains the late--time acceleration remarkably well; however, it suffers from several conceptual difficulties including the cosmological constant problem and the coincidence problem. Consequently, different phenomenological and reconstructed forms of the Hubble parameter have been proposed in the literature to explain the cosmic acceleration without explicitly assuming a cosmological constant. Such reconstruction techniques are particularly useful because they provide a model--independent description of cosmic evolution and simultaneously remain compatible with observational data.

Motivated by these considerations, in the present work we consider the following reconstructed Hubble parametrization:
\begin{equation}\label{Hz}
	H(z)=H_{0}(1+z)^{n}
	\sqrt{\frac{1+(1+z)^{-2n}}{2}},
\end{equation}
where $H_{0}$ denotes the present value of the Hubble parameter and $n$ is a free model parameter controlling the cosmic dynamics. The above parametrization possesses several physically significant features which make it suitable for describing the cosmic evolution.

First, the proposed form remains finite and positive throughout the cosmological evolution and avoids any unphysical divergence in the observable range of redshift. Moreover, the model smoothly interpolates between different expansion phases of the universe. To understand its physical behavior, let us examine its limiting cases. For large redshift values $(z\gg1)$ corresponding to the early universe, $(1+z)^{-2n}\ll1$, and therefore equation \eqref{Hz} reduces approximately to $H(z)\approx \frac{H_{0}}{\sqrt{2}}(1+z)^{n}$. This power--law behavior naturally describes the decelerating phase of the early universe dominated by matter or radiation depending upon the choice of the parameter $n$. Hence, the model successfully reproduces the standard behavior required during the structure formation epoch. On the other hand, at late cosmic times $(z\rightarrow -1)$, the second term inside the square root becomes dominant and modifies the expansion rate significantly. Consequently, the Hubble parameter evolves more slowly and approaches a nearly constant behavior, thereby generating an accelerated expansion phase. Thus, the parametrization naturally incorporates the transition from early deceleration to late acceleration without introducing an explicit cosmological constant term.

The present Hubble function is also motivated by several reconstructed and phenomenological cosmological parametrizations proposed in the literature. Mukherjee and Banerjee (\textcolor{blue}{2016}) reconstructed the effective equation of state through a generalized Hubble reconstruction approach capable of describing late--time acceleration. Similarly, different transition-type Hubble parametrizations and reconstructed dark energy models have been investigated by many authors in the context of observational cosmology (Alam et al. \textcolor{blue}{2003}; Sahni et al. \textcolor{blue}{2003}; Barboza et al. \textcolor{blue}{2008}; Riess et al. \textcolor{blue}{2004}). In particular, reconstruction techniques based on redshift-dependent Hubble parametrizations have proven to be extremely useful for obtaining observationally constrained cosmological models without assuming a specific scalar field potential or dark energy fluid from the beginning. The present parametrization follows the same philosophy and extends such approaches through a symmetric transition structure involving the factor  $\sqrt{\frac{1+(1+z)^{-2n}}{2}}$, which provides a smooth interpolation between different expansion regimes of the universe.
It is important to mention that similar reconstructed Hubble parametrizations have recently attracted considerable attention in modern cosmology due to their success in fitting cosmological observations and explaining dark energy evolution in a model--independent manner (Mukherjee and Banerjee \textcolor{blue}{2016}; Sahni et al. \textcolor{blue}{2003}; Barboza et al. \textcolor{blue}{2008}). However, the specific form considered in the present work is comparatively more flexible and analytically tractable for studying cosmological diagnostics and thermodynamic behavior within modified gravity theories. Therefore, motivated by the above physical and mathematical properties, we employ the reconstructed Hubble parametrization to investigate the cosmological dynamics of the universe in the framework of modified gravity.

In the next section, we constrain the free model parameters using recent observational datasets including the Observational Hubble Data (OHD) and the Pantheon+ Supernova compilation.

\section{Observational Constraints}
To examine the physical viability and cosmological consistency of the present model, it is essential to constrain the free parameters of the theory through observational investigations. In this regard, we estimate the model parameters \(H_{0}\), and \(n\) by employing the latest cosmological observations. In particular, the observational Hubble data (OHD) and the Pantheon+ compilation of Type Ia supernovae are utilized, since these datasets provide reliable information regarding the expansion history of the Universe through the Hubble parameter \(H(z)\) and luminosity distance measurements. The combined analysis of these datasets enables a robust statistical estimation of the cosmological parameters and facilitates a meaningful comparison of the proposed model with the standard cosmological scenario.


\subsection{Observational Hubble Data (OHD)}

In order to constrain the free parameters of the model, we make use of the recent compilation of 77 uncorrelated observational Hubble measurements \(H(z)\) distributed over the redshift interval \(0 \leq z \leq 2.36\). These data points are obtained from different observational surveys 
To ensure a statistically consistent analysis and to incorporate possible systematic uncertainties associated with the Cosmic Chronometer (CC) method, we employ the complete covariance matrix provided in Ref. Moresco et al. \textcolor{blue}{2020}. The corresponding covariance data are publicly available at
\url{https://gitlab.com/mmoresco/CCcovariance}.

The chi-square function corresponding to the OHD dataset is expressed as
\begin{equation}
	\chi^{2}_{\rm OHD}
	=
	(\mathbf{H}_{\rm obs}-\mathbf{H}_{\rm th})^{T}
	\mathbf{C}^{-1}
	(\mathbf{H}_{\rm obs}-\mathbf{H}_{\rm th}),
\end{equation}
where \(\mathbf{H}_{\rm obs}\) and \(\mathbf{H}_{\rm th}\) represent the observed and theoretical values of the Hubble parameter, respectively, while \(\mathbf{C}\) denotes the covariance matrix. The inclusion of the covariance matrix significantly improves the statistical reliability of the parameter estimation and properly accounts for correlated systematic effects in the observational data (Moresco et al. \textcolor{blue}{2020}).


\subsection{Pantheon+ Compilation}

To further investigate the cosmological behaviour of the model, we also employ the Pantheon+ sample of Type Ia supernovae (SN Ia) (Scolnic et al. \textcolor{blue}{2018}). The Pantheon+ compilation consists of 1701 light curves corresponding to 1550 spectroscopically confirmed SN Ia events spanning the redshift range
\(
0.001 \leq z \leq 2.26
\).
Because SN Ia observations serve as standardizable candles, they provide one of the most effective probes for studying the late-time accelerated expansion of the Universe.

The luminosity distance for the present cosmological model is given by
\begin{equation}
	D_{L}(z)
	=
	(1+z)
	\int_{0}^{z}
	\frac{H_{0}}{H(z^{\prime})}
	\,dz^{\prime}.
\end{equation}

Using the above expression, the theoretical distance modulus can be written as
\begin{equation}
	\mu(z)
	=
	5\log_{10}
	\left(
	\frac{D_{L}(z)}{1~{\rm Mpc}}
	\right)
	+25.
\end{equation}

It is important to note that the Pantheon+ dataset provides measurements of the distance modulus \(\mu(z)\), but these measurements are not absolutely calibrated because the absolute magnitude \(M_{B}\) of Type Ia supernovae is not intrinsically fixed within the dataset itself. Therefore, the theoretical apparent magnitude is expressed as
\begin{equation}
	m^{\rm theory}_{B,i}
	=
	\mu(z_{i})+M_{B}.
\end{equation}

Consequently, the chi-square estimator for the Pantheon+ sample takes the form
\begin{equation}
	\chi^{2}_{\rm Pantheon+}
	=
	\sum_{i,j}
	\left(
	m^{\rm data}_{B,i}
	-
	m^{\rm theory}_{B,i}
	\right)^{T}
	\mathcal{C}^{-1}_{ij}
	\left(
	m^{\rm data}_{B,j}
	-
	m^{\rm theory}_{B,j}
	\right),
\end{equation}
where \(\mathcal{C}\) denotes the covariance matrix associated with the Pantheon+ compilation, while the indices \(i\) and \(j\) run over the entire SN Ia dataset.


For the statistical estimation of the model parameters, we compare the theoretically predicted cosmological quantities with their corresponding observational values by minimizing the chi-square function. Let \(E_{\rm th}\) denote the theoretical value predicted by the model and \(E_{\rm obs}\) represent the observed quantity. The general form of the chi-square estimator is defined as
\begin{equation}
	\label{chi1}
	\chi^{2}
	=
	\sum_{i=1}^{N}
	\left[
	\frac{
		E_{\rm th}(\Phi_{c},z_{i})
		-
		E_{\rm obs}(\Phi_{l},z_{i})
	}{
		\sigma_{i}
	}
	\right]^{2},
\end{equation}
where \(\sigma_{i}\) corresponds to the standard deviation associated with the observational measurements and \(N\) denotes the total number of data points. Here,
\(
\Phi_{c}=(H_{0}, n)
\)
represents the set of cosmological model parameters, whereas \(\Phi_{l}\) denotes the nuisance parameter set associated with the corresponding observational dataset.

\begin{figure*}[h]
	\centering
	\includegraphics[scale=1.2]{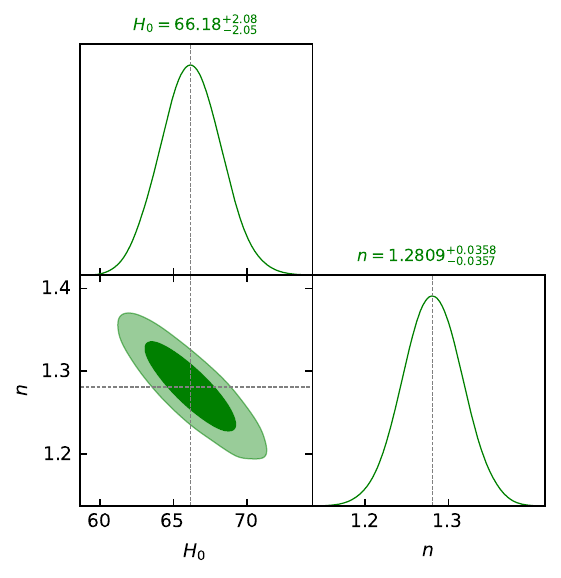}
	\caption{One-dimensional marginalized distribution and two-dimensional contours at $1\sigma$ and $2\sigma $ confidence levels of proposed model by using the combined most recent 77 observational Hubble and Pantheon+ data sets to constrain model parameters. The unit of $H_{0}$ is $\;km\;s^{-1}\;Mpc^{-1}$}.\label{OC}
\end{figure*}

The confidence contour analysis obtained from the combined observational datasets (OHD and Pantheon+) is presented in Fig.~\ref{OC}. Figure~\ref{OC} illustrates the two-dimensional likelihood contours at the \(1\sigma\) and \(2\sigma\) confidence levels derived from the 77 OHD measurements. The best-fit numerical values of the model parameters are obtained as $H_0 = 66.1839^{+2.0836}_{-2.0507}$ and $n = 1.2809^{+ 0.0358} _{- 0.0357}$.
\begin{figure*}[h]
	\centering
	\includegraphics[scale = 0.530]{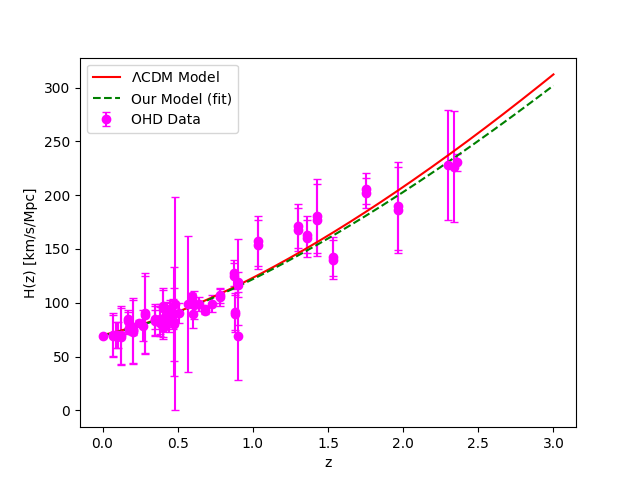}
	\includegraphics[scale = 0.530]{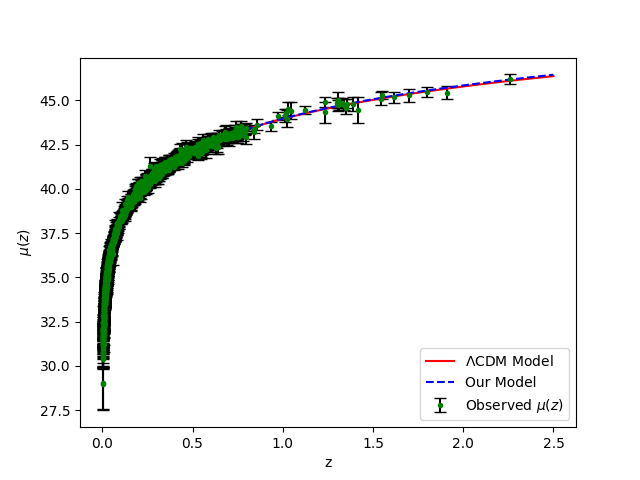}
	\caption{The left panel of above figure shows the variation of $H(z)$ of our model with redshift $z$ and its comparison with $\Lambda$CDM model while the right panel of above figure exhibits the variation of distance modulus $\mu(z)$ of our model with redshift $z$ and its comparison with $\Lambda$CDM model.}\label{FR1}
\end{figure*}
Furthermore, Fig.~\ref{FR1} demonstrates the cosmological behaviour of the proposed model through a comparison with the standard \(\Lambda\)CDM scenario. In the left panel, the evolution of the Hubble parameter \(H(z)\) with redshift \(z\) is displayed together with the associated \(1\sigma\) confidence interval. Similarly, the right panel illustrates the variation of the distance modulus \(\mu(z)\) against redshift along with its corresponding \(1\sigma\) uncertainties. From the observational analysis, it is observed that the proposed cosmological model remains in close agreement with the standard \(\Lambda\)CDM cosmology and exhibits behaviour consistent with recent observational constraints 
\section{Cosmological Dynamics}
After constraining the free model parameters through the combined observational datasets, we now investigate the physical and dynamical behaviour of the reconstructed cosmological model. In particular, the present section is devoted to the analysis of the dark energy density evolution, equation of state parameter, thermodynamic behaviour, and kinematical properties of the universe within the framework of interacting $f(Q,\mathcal{T})$ gravity.

The analysis of the dark energy density and equation of state parameter plays an important role in understanding the nature of the accelerating universe and the dynamical behaviour of the interacting dark sector. Similarly, thermodynamic investigations provide important information regarding the physical viability and stability of the cosmological model through the generalized second law of thermodynamics and entropy evolution. Furthermore, the study of kinematical parameters such as the deceleration parameter, jerk parameter, and statefinder diagnostics helps distinguish the present model from the standard $\Lambda$CDM cosmology and other dark energy scenarios.

\subsection{Evolution of Dark Energy Density and Equation of State Parameter}

In the framework of interacting $f(Q,\mathcal{T})$ gravity, the reconstructed dark energy density is obtained as
\begin{widetext}
\begin{align}
	\rho_{de}(z)
	=
	\frac{
		-\alpha\beta n H_0^2(1+z)^{2n}
		+
		\frac{\alpha(2+5\beta)}{2}
		H_0^2
		\left[(1+z)^{2n}+1\right]
	}{
		2+\beta
	}
	-
	\rho_{m0}(1+z)^3.
	\label{rho_de_z}
\end{align}
\end{widetext}

\begin{figure*}[ht]
	\centering
	\includegraphics[scale = 0.6]{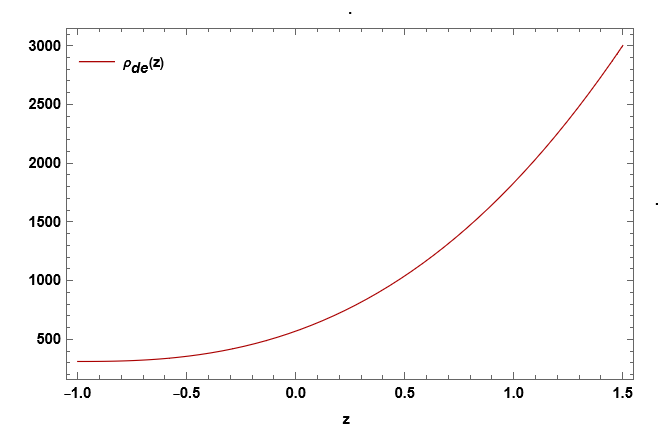}
	\includegraphics[scale = 0.6]{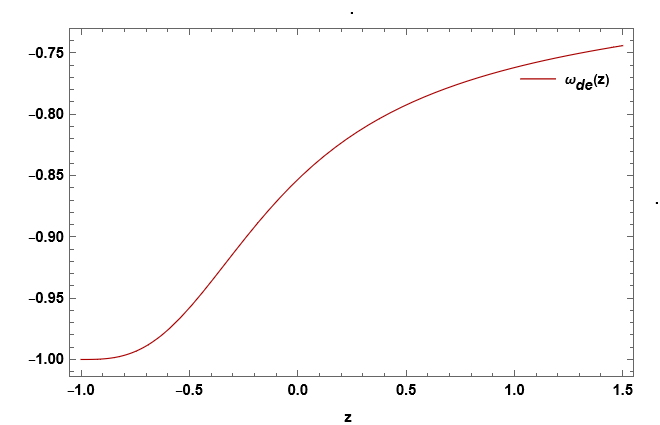}
	\caption{Evolution of dark energy density versus redshift $z$ (left panel) and the $\omega_{de}$ parameter versus redshift $z$ (right panel) for the reconstructed interacting $f(Q,\mathcal{T})$ cosmological model.}\label{den}
\end{figure*}

The above expression explicitly describes the evolution of dark energy density in terms of the redshift parameter. The graphical analysis of $\rho_{de}(z)$ is clearly seen in the figure (\ref{den}.It is observed that the behaviour of $\rho_{de}$ strongly depends upon the model parameters $\alpha$, $\beta$, and $n$. The first term in Eq.~(\ref{rho_de_z}) arises due to the modified gravity contribution associated with the nonminimal coupling between nonmetricity and matter, whereas the second term corresponds to the standard matter density evolution.

At high redshift values $(z\gg1)$, the matter contribution dominates the cosmic dynamics and the universe remains in a decelerated expansion phase. However, as the universe evolves toward lower redshift values, the modified gravity contribution gradually becomes dominant, thereby generating an effective dark energy component capable of driving the present accelerated expansion.

The positivity and smooth behaviour of the reconstructed dark energy density indicate that the present model remains physically viable throughout the cosmological evolution. Depending upon suitable choices of the model parameters, the dark energy density may either monotonically decrease or approach an asymptotically constant value at late times, thereby mimicking a cosmological constant dominated epoch. Using the reconstructed Hubble parametrization, the dark energy density parameter becomes
\begin{widetext}
\begin{align}
	\Omega_{de}(z)
	=
	\frac{2}{3H_0^2\left[(1+z)^{2n}+1\right]}
	\left[
	\frac{
		-\alpha\beta n H_0^2(1+z)^{2n}
		+
		\frac{\alpha(2+5\beta)}{2}
		H_0^2
		\left[(1+z)^{2n}+1\right]
	}{
		2+\beta
	}
	-
	\rho_{m0}(1+z)^3
	\right].
	\label{omega_density}
\end{align}
\end{widetext}
Substituting Eq.~(\ref{omega_density}) into Eq.~(\ref{omegaflat}), the reconstructed equation of state parameter becomes
\begin{widetext}
\begin{align}
	\omega_{de}(z)=-\frac{3H_0^2\left[(1+z)^{2n}+1\right]\left[\displaystyle\frac{-\alpha\beta n H_0^2(1+z)^{2n}+\frac{\alpha(2+5\beta)}{2}H_0^2\left[(1+z)^{2n}+1\right]}{2+\beta}-\rho_{m0}(1+z)^3\right]+6\xi H_0^2\left[(1+z)^{2n}+1\right]
	}{\left[6H_0^2\left[(1+z)^{2n}+1\right]	-2\left(\displaystyle\frac{-\alpha\beta n H_0^2(1+z)^{2n}+\frac{\alpha(2+5\beta)}{2}H_0^2\left[(1+z)^{2n}+1\right]}{2+\beta}-\rho_{m0}(1+z)^3\right)\right]}.
\end{align}
\end{widetext}

The evolution of $\omega_{de}$ provides important information regarding the dynamical nature of dark energy. In particular, the present model allows various possible cosmological behaviours depending upon the values of the coupling parameter $\xi$ and the reconstructed parameter $n$. The graphical analysis of $\omega_{de}(z)$ is clearly shown in the figure (\ref{den}). For suitable choices of parameters, the equation of state parameter may remain in the quintessence region $(\omega_{de}>-1)$ and smoothly approach the cosmological constant boundary $(\omega_{de}=-1)$ at late times.
Moreover, the interaction between dark matter and dark energy significantly influences the dynamical behaviour of the equation of state parameter. An increase in the interaction strength modifies the late-time evolution of $\omega_{de}$ and may generate a smooth transition across the phantom divide line. Therefore, the interacting framework in $f(Q,\mathcal{T})$ gravity provides a flexible and observationally consistent description of the present accelerated universe.

\subsection{Thermodynamic Behaviour of the Model}
The thermodynamic behaviour of the universe plays a fundamental role in understanding the physical viability and stability of cosmological models. In modified theories of gravity, particularly in the framework of $f(Q,\mathcal{T})$ gravity, the gravitational field equations possess an explicit coupling between geometry and matter, which leads to important modifications in the standard thermodynamic description of the universe.

In the present work, we investigate the thermodynamic properties of the reconstructed cosmological model by considering the apparent horizon of the spatially flat FRW universe as the thermodynamic boundary. Since the apparent horizon is associated with gravitational entropy and Hawking temperature, it provides a natural framework for studying the generalized laws of thermodynamics in cosmology.

For the flat FRW space--time, the radius of the apparent horizon is given by $R_A=\frac{1}{H}$. Using the reconstructed Hubble parametrization, the apparent horizon radius becomes
\begin{align}
	R_A(z)
	=
	\frac{
		\sqrt{2}
	}{
		H_0\sqrt{(1+z)^{2n}+1}
	}.
\end{align}

\begin{figure*}[ht]
	\centering
	\includegraphics[scale = 0.19]{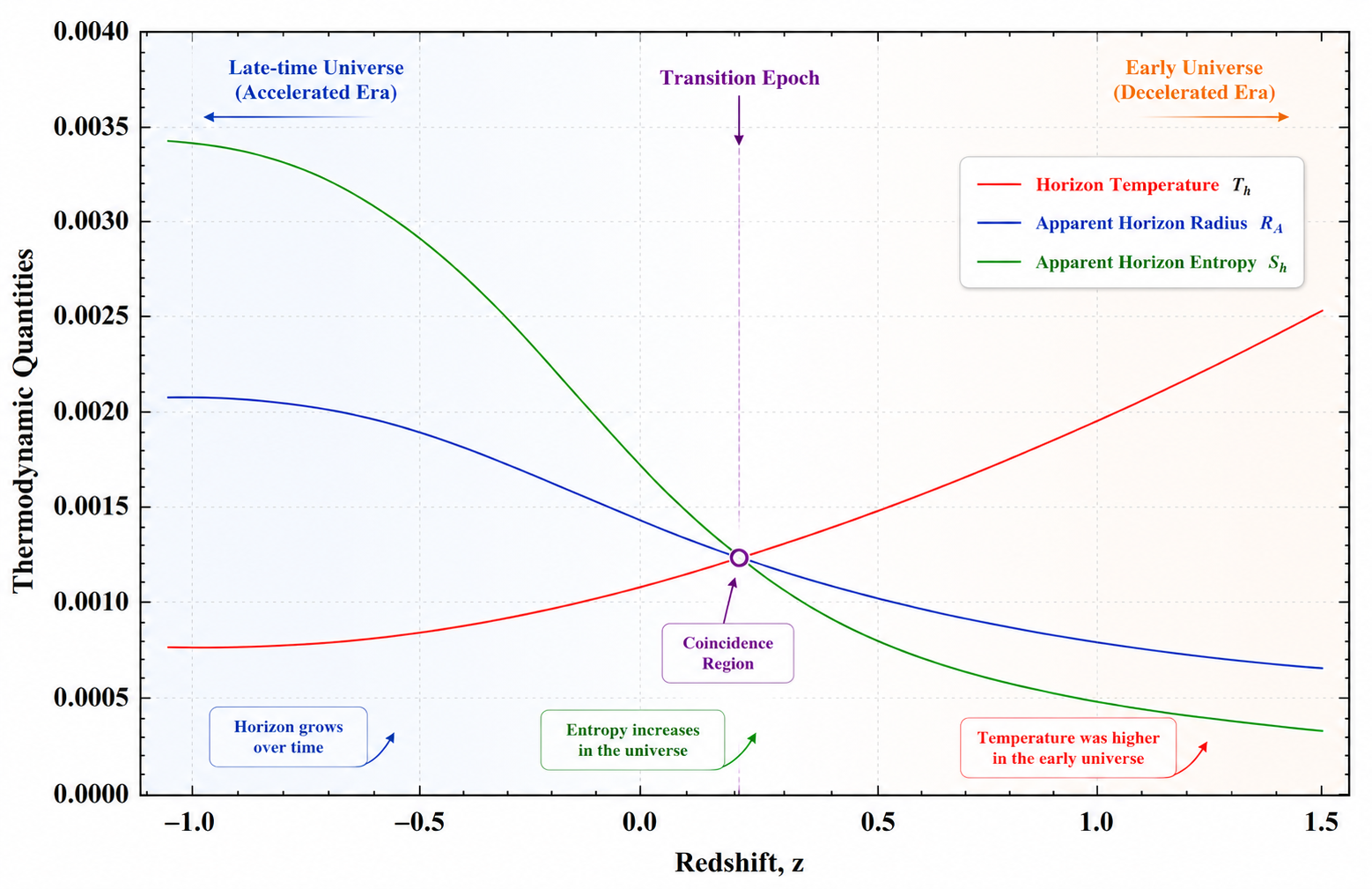}
	\caption{Evolution of the thermodynamic quantities namely the horizon temperature $T_h$, the apparent horizon radius $R_A$, and the apparent horizon entropy $S_h$ versus redshift $z$ for the reconstructed interacting $f(Q,\mathcal{T})$ cosmological model.}\label{Thermo}
\end{figure*}

The Hawking temperature associated with the apparent horizon is expressed as $T_h=\frac{1}{2\pi R_A}=	\frac{H}{2\pi}$. Therefore, the horizon temperature takes the form
\begin{align}
	T_h(z)=\frac{H_0(1+z)^n}{2\pi}\sqrt{\frac{1+(1+z)^{-2n}}{2}}.
\end{align}
In modified gravity theories, the entropy-area relation receives corrections due to the modification of the gravitational action. In the framework of $f(Q,\mathcal{T})$ gravity, the horizon entropy is defined as $S_h	=\frac{Af_Q}{4G}$, where $A=4\pi R_A^2$ denotes the area of the apparent horizon. For the linear form $f(Q,\mathcal{T})$, one has $f_Q=\alpha$. Hence, the modified horizon entropy becomes $S_h=\frac{\alpha A}{4G}$. Using the natural units $(8\pi G=1)$ and the apparent horizon area, one obtains $S_h=8\pi^2\alpha R_A^2$. Consequently,
\begin{align}
	S_h(z)
	=
	\frac{
		16\pi^2\alpha
	}{
		H_0^2\left[(1+z)^{2n}+1\right]
	}.
\end{align}
Figure~(\ref{Thermo}) illustrates the evolution of the thermodynamic quantities namely the apparent horizon radius $R_A$, the horizon temperature $T_h$, and the horizon entropy $S_h$ as functions of the redshift parameter $z$. The behaviour of these quantities provides important information regarding the thermodynamic nature and physical viability of the reconstructed interacting $f(Q,\mathcal{T})$ cosmological model.

It is observed from the figure that the apparent horizon radius decreases smoothly with increasing redshift. This behaviour indicates that the horizon radius was comparatively small during the early cosmic epoch when the expansion rate of the universe was very high. On the other hand, toward the late-time accelerated era $(z\rightarrow -1)$, the apparent horizon radius increases continuously due to the gradual decrease of the Hubble expansion rate. The increasing horizon radius at late times reflects the continuous growth of the causal boundary of the universe during cosmic expansion.

The horizon temperature exhibits the opposite behaviour. It increases with increasing redshift and attains larger values in the early universe. This behaviour is physically expected because the Hubble parameter remains large during the primordial epoch, resulting in a high horizon temperature. As the universe evolves toward the late accelerated phase, the temperature gradually decreases, indicating the cooling behaviour of the expanding universe. Furthermore, the positivity of the horizon temperature throughout the cosmic evolution confirms the thermodynamic consistency of the cosmological model.

The apparent horizon entropy also decreases with increasing redshift and becomes larger during the late-time evolution of the universe. Since entropy is directly proportional to the area of the apparent horizon, the increase in entropy at lower redshift is a consequence of the growth of the horizon radius during cosmic expansion. This behaviour indicates that the total disorder of the universe continuously increases with time, which is consistent with the generalized second law of thermodynamics.

Moreover, all the thermodynamic quantities evolve smoothly without exhibiting any divergence or unphysical behaviour over the entire redshift range. The regular and positive behaviour of the horizon temperature and entropy suggests that the present interacting $f(Q,\mathcal{T})$ gravity model remains thermodynamically stable and physically viable throughout the cosmic evolution.

Therefore, the thermodynamic analysis demonstrates that the reconstructed cosmological model provides a physically acceptable description of the late-time accelerated universe and remains consistent with the fundamental laws of thermodynamics.

Hence, the obtained thermodynamic behaviour confirms that the present interacting $f(Q,\mathcal{T})$ cosmological model is thermodynamically admissible, stable, and capable of describing the observed accelerated expansion of the universe in a physically consistent manner.
\subsection{Kinematical Properties of the Model}

In order to further investigate the cosmic evolution and distinguish the present model from other dark energy scenarios, we now study several important kinematical parameters associated with the expansion dynamics of the universe.
\subsubsection{The deceleration parameter} 
\begin{figure*}[ht]
	\centering
	\includegraphics[scale = 0.60]{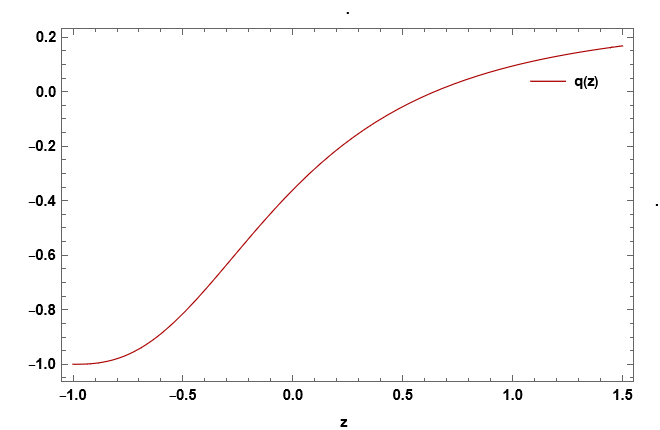}
	\caption{The above figure shows the variation of $q$ with redshift $z$.}\label{q}
\end{figure*}

The deceleration parameter is defined as $q=-1-\frac{\dot H}{H^2}$. Using the reconstructed Hubble parametrization, the deceleration parameter becomes
\begin{align}
	q(z)=-1+\frac{n(1+z)^{2n}}{(1+z)^{2n}+1}.
\end{align}
This expression describes the dynamical evolution of the expansion rate of the Universe in terms of the redshift parameter \(z\). The graphical behavior of \(q(z)\) in figure (\ref{q}) clearly demonstrates a smooth transition of the Universe from a decelerated phase in the past to an accelerated phase at the present epoch and in the future. For large positive redshift (\(z\gg0\)), the term \((1+z)^{2n}\) dominates over unity and therefore the deceleration parameter approaches $q(z)\approx -1+n$. For the chosen model parameter \(n=1.280\), one obtains $q(z)\approx 0.28$, which is positive. This indicates that the early Universe was in a decelerated expansion phase, consistent with the standard matter-dominated cosmological era. The positive value of \(q\) signifies that gravitational attraction dominated the cosmic dynamics at earlier times.

As the Universe evolves toward lower redshift, the deceleration parameter decreases continuously and eventually crosses the line \(q=0\). The transition redshift \(z_t\) is obtained from the condition $q(z_t)=0$. Using the given expression, $-1+\frac{n(1+z_t)^{2n}}{(1+z_t)^{2n}+1}=0$, which yields $(1+z_t)^{2n}=\frac{1}{n-1}$. Hence, $z_t=\left(\frac{1}{n-1}\right)^{\frac{1}{2n}}-1$. Substituting \(n=1.280\), the transition redshift becomes approximately
\begin{align}
	z_t \approx 0.65.
\end{align}
Thus, the model predicts that the Universe changed from decelerated expansion to accelerated expansion around \(z\simeq0.65\), which lies well within the observationally accepted range obtained from recent cosmological datasets. The graphical profile also confirms this transition point where the curve intersects the \(q=0\) axis. At the present epoch (\(z=0\)), the deceleration parameter becomes $q_0=-1+\frac{n}{2}$. For \(n=1.280\), $q_0 \approx -0.36$. 
The negative present value of the deceleration parameter indicates that the Universe is currently undergoing accelerated expansion. This result is in good agreement with observational evidence from Type Ia Supernovae, BAO, and CMB measurements, which strongly support the existence of late-time cosmic acceleration.

In the future evolution of the Universe (\(z\to -1\)), the quantity \((1+z)^{2n}\to0\), and therefore $q(z)\to -1$. This asymptotic behavior corresponds to a de Sitter type accelerated phase, where the expansion becomes exponentially accelerating. Hence, the model naturally evolves toward a dark-energy-dominated epoch in the far future.
Overall, the behavior of the deceleration parameter reveals that the proposed cosmological model successfully explains the complete cosmic evolution sequence: an initially decelerating Universe, a smooth transition phase near \(z\approx0.65\), the presently accelerating epoch with \(q_0\approx-0.36\), and finally a de Sitter accelerated future characterized by \(q\to-1\). The smooth and monotonic nature of the curve further indicates the absence of any sudden cosmological singularity or instability during the cosmic evolution.
\subsubsection{The statefinder diagnostic pair}
To further investigate the dynamical nature of the present cosmological model, we analyze the statefinder diagnostic pair \((r,s)\), which is defined as
\begin{align}
	r=q+2q^2+\frac{\dot{q}}{H},
	\qquad
	s=\frac{r-1}{3\left(q-\frac12\right)}.
\end{align}

\begin{figure*}[ht]
	\centering
	\includegraphics[scale = 0.60]{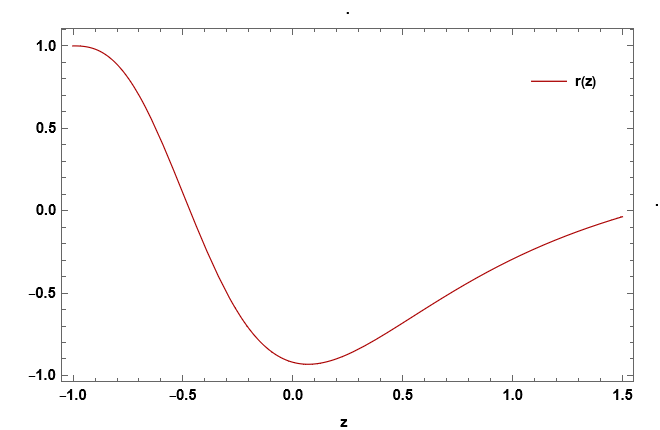}
	\includegraphics[scale = 0.60]{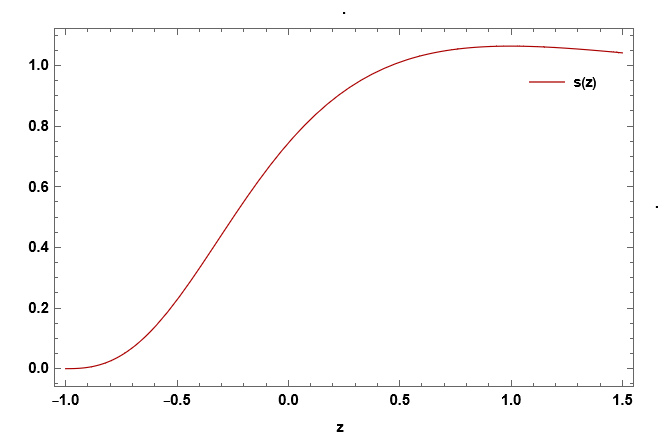}
	\caption{The left panel of above figure shows the variation of $r(z)$ with redshift $z$ while the right panel of above figure exhibits the variation of $s(z)$ of our model with redshift $z$.}\label{rs}
\end{figure*}

The statefinder parameters provide an efficient geometrical diagnostic to distinguish different dark energy models including quintessence, Chaplygin gas, holographic dark energy, and modified gravity models. In particular, trajectories approaching the fixed point $(r,s)=(1,0)$ correspond to the standard $\Lambda$CDM cosmology and describe the evolutionary behavior of the Universe beyond the Hubble and deceleration parameters. Using the obtained deceleration parameter, the corresponding evolutionary trajectories of \(r(z)\) and \(s(z)\) are illustrated in Fig. (\ref{rs}).
The graphical behavior of the left panel shows that the parameter \(r(z)\) evolves smoothly from positive values in the early Universe to negative values at intermediate redshift and finally approaches to 1 in the late-time epoch. On the other hand, the right panel demonstrates that \(s(z)\) increases monotonically from nearly zero in the future epoch toward values close to unity during the past evolution of the Universe.

For large redshift (\(z\gg0\)), corresponding to the early decelerating phase of the Universe, the deceleration parameter approaches $q(z)\approx0.28$. Substituting this behavior into the statefinder expressions yields approximately $r(z), s(z) \ge 0$. Thus, in the high-redshift regime the model behaves similarly to a standard matter-dominated cosmological phase. The positive value of \(r\) and large value of \(s\) indicate that the Universe was initially dominated by decelerating matter dynamics. As the Universe evolves toward lower redshift, both statefinder parameters exhibit significant variation. The parameter \(r(z)\) decreases continuously and crosses the zero line around the transition region where the cosmic expansion changes from deceleration to acceleration. Simultaneously, the parameter \(s(z)\) increases rapidly, indicating the gradual dominance of dark energy over matter components. The transition from decelerated to accelerated expansion occurs near the redshift $z_t \approx 0.65$, which is consistent with the behavior of the deceleration parameter discussed previously. Around this epoch, the statefinder pair exhibits a clear dynamical transition, demonstrating the departure of the model from the matter-dominated phase toward an accelerating dark-energy-dominated phase.

At the present epoch (\(z=0\)), where the deceleration parameter is $q_0\approx-0.36$, the corresponding statefinder parameters are approximately obtained as $r_0\approx -0.92,  s_0\approx0.76$. The negative present value of \(r_0\) reflects the strong influence of dark energy in the current accelerating phase, whereas the positive value of \(s_0\) indicates that the model dynamically deviates from the standard \(\Lambda\)CDM cosmology while still remaining compatible with late-time accelerated expansion. In the future evolution of the Universe (\(z\rightarrow-1\)), the quantity \((1+z)^{2n}\rightarrow0\), which leads to $q(z)\rightarrow-1$. Consequently, the statefinder parameters asymptotically approach $r(z)\rightarrow 1, s(z)\rightarrow 0$. This behavior implies that the Universe eventually evolves toward a stable accelerated phase dominated entirely by dark energy and behave as $\Lambda$CDM model. The future convergence of the statefinder pair toward small values signifies the stabilization of cosmic expansion in the asymptotic epoch.

Overall, the evolutionary trajectories of the statefinder parameters \((r,s)\) clearly reveal the complete cosmic history of the model. The Universe begins in a matter-dominated decelerating phase characterized by positive \(r\) and large \(s\), undergoes a smooth transition near \(z\simeq0.65\), and finally evolves toward a stable accelerated epoch in the future. The smooth and continuous evolution of both parameters also confirms the absence of any abrupt cosmological instability or singular behavior in the considered cosmological framework.

\subsubsection{The Om diagnostic}
The evolution of the \(Om(z)\) diagnostic parameter provides an important geometrical tool to distinguish the present cosmological model from the standard \(\Lambda\)CDM scenario defined as $Om(z)=\frac{\left(\frac{H(z)}{H_0}\right)^2-1}{(1+z)^3-1}$. For the obtained Hubble parametrization, the \(Om(z)\) diagnostic is expressed as
\begin{align}
	Om(z)=
	\frac{
		\frac{1}{2}(1+z)^{2n}\left[(1+z)^{-2n}+1\right]-1
	}{
		(1+z)^3-1
	}.
\end{align}

The \(Om(z)\) diagnostic depends only on the Hubble parameter and redshift, making it a model-independent and efficient method for investigating the nature of dark energy. In the standard \(\Lambda\)CDM cosmology, the \(Om(z)\) parameter remains constant and equals the present matter density parameter \(\Omega_{m0}\). Therefore, any deviation from a constant behavior indicates the existence of dynamical dark energy or modified gravity effects. For the present model, the behavior of \(Om(z)\) strongly depends on the model parameter \(n\). At high redshift (\(z\gg1\)), the denominator \((1+z)^3-1\) becomes dominant, while the numerator evolves according to the power-law contribution governed by \(n\). This reflects the influence of the modified cosmic expansion history during the matter-dominated epoch. Near the present epoch (\(z=0\)), the \(Om(z)\) parameter approaches a finite value that characterizes the current matter contribution of the Universe. The smooth evolution of the diagnostic indicates a stable cosmological transition from earlier decelerated expansion to the present accelerated phase. Furthermore, the slope of the \(Om(z)\) trajectory provides information about the nature of dark energy. A positive slope of \(Om(z)\) generally corresponds to phantom-like behavior \((\omega<-1)\), whereas a negative slope indicates quintessence-like behavior \((\omega>-1)\). The obtained evolution of the diagnostic therefore serves as an efficient indicator for examining the dynamical properties of the dark energy sector within the considered cosmological framework.

In the far future limit (\(z\rightarrow -1\)), the denominator tends toward a finite asymptotic behavior, and the evolution of \(Om(z)\) reflects the late-time dominance of dark energy. The overall behavior of the diagnostic demonstrates that the proposed model successfully reproduces the observed cosmic acceleration while allowing possible deviations from the standard \(\Lambda\)CDM cosmology.

\section{CONCLUSION}

In this research, we have explored the cosmological implications of the $f(Q,\mathcal{T})$ gravity framework, specifically focusing on the coupling between the non-metricity scalar $Q$ and the trace of the energy-momentum tensor $\mathcal{T}$. 
We investigated the evolution of a flat FRW universe. The major highlights and detailed findings of this study are as follows:

\begin{itemize}
	\item \textbf{Formulation of the field equations:} We formulated the field equations for an FRW universe in $f(Q,\mathcal{T})$ gravity using the functional form $f(Q,\mathcal{T}) = \alpha Q + \beta \mathcal{T}$. This choice allows for a direct coupling between the geometry and the matter sector, which significantly influences the cosmic evolution.
	
	\item \textbf{$H(z)$ parametrization:} By employing the $H(z)$ parametrization, we derived the exact solutions for the scale factor and other cosmological parameters. The model successfully depicts a smooth transition from the radiation-dominated era to the current dark energy-dominated accelerated phase.
	
	\item \textbf{Observational Constraints:} Using the $H(z)$ and Pantheon+ datasets, we performed a statistical analysis to determine the best-fit values for the model parameters. The constraint values were found to be $H_0 \approx 66.18^{+2.08} _{-2.05}$ km/s/Mpc and the power-law index $n \approx 1.2809^{+0.0358}_{-0.0357}$. For the modified gravity parameters, we utilized $\alpha = 0.1$ and $\beta = 0.01$, which ensure the model's alignment with current observational bounds.
	
	\item \textbf{Kinematical Diagnostics ($q$ and $\omega_{de}$):} The deceleration parameter $q$ exhibits a smooth transition from a positive value ($q > 0$) in the past to a negative value ($q < 0$) at present, with a transition redshift $z_t \approx 0.65$. This confirms the shift from a decelerated to an accelerated expansion phase. Furthermore, the dark energy equation of state parameter $\omega_{de}(z)$ starts from the quintessence region and evolves towards the phantom divide line ($\omega_{de} = -1$) as $z \to -1$, effectively mimicking the behavior of dynamical dark energy like $\Lambda$CDM model.
	
	\item \textbf{Statefinder Diagnostics ($r, s$):} To distinguish our $f(Q,\mathcal{T})$ model from the standard $\Lambda$CDM model, we analyzed the statefinder parameters $\{r, s\}$. The trajectory in the $r-s$ plane shows a distinct deviation from the $\Lambda$CDM fixed point $\{1, 0\}$, eventually converging towards it in the future. This confirms that while the model differs from $\Lambda$CDM at high redshifts, it serves as a viable dark energy candidate in the late-time limit.
	
	\item \textbf{Thermodynamic properties:} The apparent horizon entropy was found to be an increasing function of time (decreasing with redshift), which is directly proportional to the growth of the horizon area. The smooth evolution of all thermodynamic quantities without any divergence indicates that the $f(Q,\mathcal{T})$ gravity model is thermodynamically stable and physically admissible.
\end{itemize}

Overall, our investigation demonstrates that $f(Q,\mathcal{T})$ gravity provides a robust and stable theoretical framework. The integration of thermodynamic consistency with observational constraints makes it a promising alternative for understanding the dark sector of the universe.
\section*{Declaration of competing interest}
The authors declare that they have no known competing financial interests or personal relationships that could have appeared to influence the work reported in this paper.

\section*{Data availability}
No data was used for the research described in the article.

\section*{Acknowledgments}
S. H. Shekh and Pankaj Kumar gratefully acknowledge the Inter-University Centre for Astronomy and Astrophysics (IUCAA), Pune, India, for their support through the Visiting Associateship program. 

\section*{Declaration of funding} The authors declare that no funds, grants, or other support were received.
		
\end{document}